\documentclass[aps,prc,preprint,superscriptaddress,amsmath,amssymb,nofootinbib]{revtex4-2}
\usepackage{fontspec,amsmath,amssymb,graphicx,booktabs,adjustbox,hyperref}
\hypersetup{colorlinks=true,linkcolor=black,citecolor=black,urlcolor=blue}
\begin{document}
\title{Spin correlations of primary fission fragments: zero-point transverse motion and the angular-momentum reservoir}
\author{D. E. Lyubashevsky}
\email{dima_physicist@mail.ru}
\affiliation{Department of Nuclear Physics, Voronezh State University, Voronezh 394006, Russian Federation}
\author{J. D. Shcherbina}
\affiliation{Department of Nuclear Physics, Voronezh State University, Voronezh 394006, Russian Federation}
\author{A. C. Perez Moreno}
\affiliation{Department of Nuclear Physics, Voronezh State University, Voronezh 394006, Russian Federation}
\affiliation{Frank Laboratory of Neutron Physics, Joint Institute for Nuclear Research, Dubna 141980, Russian Federation}
\author{S. V. Klyuchnikov}
\affiliation{Department of Nuclear Physics, Voronezh State University, Voronezh 394006, Russian Federation}
\author{D. A. Yakunin}
\affiliation{Department of Nuclear Physics, Voronezh State University, Voronezh 394006, Russian Federation}
\begin{abstract}
\textbf{Background.} Weak correlations between fragment-spin magnitudes do not identify their formation mechanism. A large moment of inertia of relative motion suppresses the kinematic coupling imposed by angular-momentum conservation, but a cold quantum state can retain potential coupling. \textbf{Purpose.} To determine the sign and magnitude of correlations in cold axial scission configurations and distinguish the role of the angular-momentum reservoir from that of the restoring interaction. \textbf{Methods.} For ²³³Th, ²³⁶U, ²³⁹U, ²⁴⁰Pu, and ²⁵²Cf, the angular Hessian and inertias of reconstructed configurations determine the coupled harmonic ground state. We analyze 839 baseline and 452 alternative energy-constrained solutions and compare quantum and planar classical canonical covariances on identical inputs. \textbf{Results.} Mean baseline mode energies are 1.10--1.17 and 2.01--2.14 MeV. The sign of component correlation follows from the difference between normalized mixed stiffness and kinematic coupling. Although the wriggling width exceeds the bending width by factors of 5.37--5.68, their contributions to the spin scalar product largely cancel. The ground-state obtuse-angle fractions are 41.6--43.0\%, whereas the same-inertia classical limit gives 55.1--55.2\%. With fixed stiffness, an infinite relative-motion inertia leaves positive ground-state correlations. Mean pairwise magnitude correlations are only 0.019--0.027. \textbf{Conclusions.} The angular-momentum reservoir alone does not determine the spin correlation of the cold state. The contrasting signs reveal the competition between potential and kinematic coupling, furnishing a joint test of absolute spins and spin orientation within the harmonic model.
\end{abstract}
\maketitle
\section{Introduction}\label{introduction}

The origin of fission-fragment spins is closely connected with the stage of the evolution at which the observed angular momenta are generated. Wilson et al.~\cite{ref2} measured substantial average residual-fragment spins and found little dependence on the spectroscopic selection of the partner's spin. They inferred that a considerable part of the angular momentum might develop after scission. Weak correlations of magnitudes, however, do not directly establish either the independence of spin directions or the time of spin generation. Randrup and Vogt \cite{ref3} found weak magnitude correlations in a model of rotational modes excited before scission. The microscopic calculations \cite{ref4,ref5,ref6} underline the significance of the joint spin distribution and longitudinal spin projections. A single magnitude correlation therefore cannot discriminate between the proposed mechanisms.

In our earlier work \cite{ref1}, spins were generated by zero-point wriggling and bending oscillations of a cold fissioning nucleus near scission. Magnitude correlations and angular distributions were evaluated for three fissioning systems. The two mode energies were set to 2.5 and 0.9 MeV, respectively, and the inertias were estimated in the superfluid-nucleus approximation. The collective distributions were treated as independent. Small correlations demonstrated the compatibility of weak magnitude correlations with a prescission mechanism. Their stability against changes in the underlying collective parameters remained to be established.

Here we retain the hypothesis that transverse collective degrees of freedom are in their ground state, while dispensing with a universal prescription for their energies. Restoring forces follow from the angular dependence of an effective nuclear and Coulomb potential; inertias follow from reconstructed shapes. Mode energies and joint momentum fluctuations are calculated in the coupled problem. The explicit potential coupling permits an examination of the angular-momentum reservoir proposed by Randrup \cite{ref17} in a different, cold quantum regime.

Zero-point angular motion as a source of spins was discussed by Nix and Świątecki \cite{ref8}, our group \cite{ref1,ref12,ref16}, and others. Shneidman et al.~\cite{ref14} studied angular states of two touching fragments in a quantum two-center model, and Ref. \cite{ref15} investigated connections between angular motion, kinetic energy, and evaporation. The use of quantum angular oscillations is thus not in itself novel. Randrup \cite{ref17} demonstrated that different assumptions about scission geometry lead to different angular distributions. The present contribution is an explicit sign criterion for the coupled ground state, applied systematically to five systems, together with a controlled comparison of quantum and classical covariances at the same inertias. We also separate the within-channel and between-channel covariances. This identifies which part of the statistical reservoir interpretation survives in the cold model.

We study ²³²Th(n,f), ²³⁵U(n,f), ²³⁸U(n,f), ²³⁹Pu(n,f), and ²⁵²Cf(sf), forming ²³³Th, ²³⁶U, ²³⁹U, ²⁴⁰Pu, and ²⁵²Cf. Primary spins for ²³⁶U and ²⁴⁰Pu are compared with microscopic AMPLIFI distributions \cite{ref7}; for the other three systems the residual spins of Wilson et al.~\cite{ref2} are used with an explicit evaporation prescription. Spin orientations are compared with FREYA \cite{ref3} and the joint quantum projection of Scamps et al.~\cite{ref6}. These tests address distinct observables and cannot be replaced by agreement with a single average spin.

\section{Cold axial prescission configurations}\label{cold-axial-prescission-configurations}

ChatGPT (OpenAI) assisted with the development and checking of the analysis scripts and mathematical expressions, and with plotting code. Numerical results are obtained from the specified model inputs and numerical routines (including the identified Monte Carlo checks), not from language-model-generated numerical data. The accompanying source code and input tables specify the computational procedure; no new microscopic event data were generated by the language model.

\subsection{Reconstruction of shapes and the angular potential}\label{reconstruction-of-shapes-and-the-angular-potential}

The principal axes of the axially deformed prefragments are aligned with the separation axis in the reference configuration. Coldness refers here to the ground state of the transverse collective degrees of freedom under consideration, not to vanishing total excitation energy; energy stored in nonequilibrium deformation can be released during relaxation. Axial symmetry makes the potential independent of a rotation of either spheroid about its own longitudinal axis. The corresponding harmonic twisting frequency vanishes, and the selected collective ground-state sector with zero longitudinal projection contains no twisting excitation. This restriction applies to the present reduction, not to arbitrary nuclei with intrinsic projections or broken axial symmetry \cite{ref8,ref6,ref19}.

For each prefragment we use a deformation-energy prescription related to the energy available for evaporation \cite{ref13}. Here \(U_i=7\nu_i+3\) is in MeV, with \(\nu_i\) the average multiplicity or an integer neutron count for a conditional channel. In this limiting prescription \(U_i\) is assigned to nonequilibrium deformation. Denoting the equilibrium and scission quadrupole deformations by \(\beta_{\mathrm{eq}}\) and \(\beta_{\mathrm{sc}}\), respectively, we obtain

\begin{equation}
F(\beta_{\mathrm{sc}})-F(\beta_{\mathrm{eq}})=U_i,\qquad U_i=7\nu_i+3.\tag{1}
\end{equation}

\begin{equation}
F(\beta)=c_2\beta^2+c_3\beta^3,\quad x=\frac{Z^2}{45 A},\tag{2}
\end{equation}

\[c_2=16 A^{2/3}\,0.4(1-x)(2/3)^2,\]

\[c_3=-16 A^{2/3}\,0.0381(1-2x)(2/3)^3.\]

Equilibrium deformations are taken from FRDM(2012) \cite{ref9}; 32 missing entries were supplemented from the original FRDM tables. Equation (1) is solved on the positive branch in the interval 0--1.5, retaining the sign of negative \(\beta_{\mathrm{eq}}\) in the cubic term. The elongated branch corresponds to the assumed prescission shape. Equation (1) is a model prescription, not a measurement of deformation energy. Separating intrinsic excitation, deformation, and zero-point energy requires additional information.

Each prefragment is approximated by a spheroid with transverse and longitudinal semiaxes \(b,c\). With \(s=\sqrt{5/(16\pi)}\), \(b_0=1-s\beta\), \(c_0=1+2s\beta\), and \(R=r_0A^{1/3}\), we conserve the volume of the corresponding sharp surface:

\begin{equation}
b=\frac{Rb_0}{(b_0^2c_0)^{1/3}},\qquad c=\frac{Rc_0}{(b_0^2c_0)^{1/3}}.\tag{3}
\end{equation}

A diffuse density is used for the nuclear potential. In the spheroid frame, \(q=c/b\), \(p=(x^2+y^2+z^2/q^2)^{1/2}\), and \(g=(\nabla p\cdot\nabla p)^{1/2}\). The radius scale \(r_0\) is fixed separately for each prefragment by normalizing the density:

\begin{equation}
\rho_N(\mathbf r)=\frac{\rho_0}{1+\exp[(p-b)/(ag)]},\qquad \int\rho_N(\mathbf r)\,d^3r=A.\tag{4}
\end{equation}

We use \(\rho_0=0.16\) fm⁻³ and \(a=0.55\) fm. Gaussian quadratures in the radial and polar variables give a maximum absolute normalization error of \(1.96\times10^{-6}\) nucleons across the 1291 available solutions. The inertia is not evaluated from the diffuse density itself but from the explicitly specified rigid sharp-surface approximation below.

At fixed center separation \(D\), the prefragments are rotated by \(\theta_1,\theta_2\) about a common transverse axis. The nuclear potential is the difference of the energy of the combined and separated densities in the local isoscalar approximation:

\begin{equation}
V_N=\int d^3r\left\{\frac{3t_0}{4}\rho_1\rho_2+\frac{t_3}{16}\left[(\rho_1+\rho_2)^{2+\eta}-\rho_1^{2+\eta}-\rho_2^{2+\eta}\right]\right\}.\tag{5}
\end{equation}

The parameters are \(t_0=-2645\) MeV fm³, \(t_3=15595\) MeV fm\(^{3+3\eta}\), and \(\eta=1/6\). This is the local part of SkM*, not the full self-consistent energy-density functional: gradient, kinetic, isovector, pairing, and spin-orbit terms are not evaluated. The Coulomb component is calculated for two uniformly charged spheroids with charge densities normalized to \(Z_1\) and \(Z_2\):

\begin{equation}
V_C=e^2\int d^3r_1d^3r_2\,\frac{\rho_{p1}(\mathbf r_1)\rho_{p2}(\mathbf r_2)}{[(\mathbf r_1-\mathbf r_2)\cdot(\mathbf r_1-\mathbf r_2)]^{1/2}},\qquad e^2=1.4399764\;\mathrm{MeV\,fm}.\tag{6}
\end{equation}

The effective interaction \(V=V_N+V_C\) is expanded about the axial configuration. For either transverse direction, the second derivatives yield a symmetric Hessian \(K\):

\begin{equation}
V(D,\boldsymbol\theta)=V(D,0)+\tfrac12\boldsymbol\theta^{\mathsf T}K\boldsymbol\theta+O((\boldsymbol\theta^{\mathsf T}\boldsymbol\theta)^2),\qquad K_{ij}=\left(\frac{\partial^2 V}{\partial\theta_i\partial\theta_j}\right)_{\boldsymbol\theta=0}.\tag{7}
\end{equation}

Reflection symmetry eliminates the linear terms and all terms odd under simultaneous reversal of the angles. The diagonal elements are obtained from \(2[V(h,0)-V(0,0)]/h^2\) and the analogous expression for the second angle; the off-diagonal element follows from \([V(h,h)-V(h,-h)]/(2h^2)\). Nine nodes are used, with \(h=0.02\) rad and Richardson extrapolation \(K=[4K(h)-K(2h)]/3\). The nuclear integral covers −13 to +13 fm in each coordinate, with a nominal step of 0.30 fm; reflection symmetry reduces the number of grid points. The Coulomb quadrature has 10×10×12 nodes per spheroid. Control runs use a step of 0.35 fm and an 8×8×10 quadrature. These settings specify numerical precision, not the physical uncertainty of the interaction.

\begin{figure}[htbp]
\centering
\includegraphics[width=0.98\linewidth,height=.65\textheight,keepaspectratio]{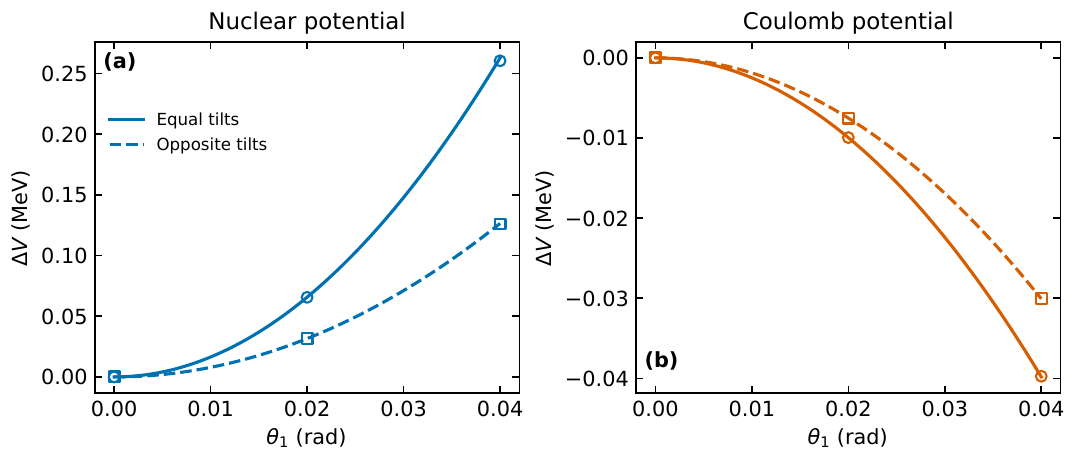}
\caption{Nuclear (a) and Coulomb (b) potential increments for ⁹⁸Sr--¹³⁸Xe. Markers are integration nodes; lines are quadratic approximations. The second tilt equals or opposes the first. Plotting code was prepared with assistance from ChatGPT (OpenAI); curves and markers are based on the stated numerical inputs.}
\end{figure}

As an example, consider ⁹⁸Sr--¹³⁸Xe in ²³⁶U fission. For \(\nu_1=1.504\) and \(\nu_2=1.006\), \(U_1=13.528\) and \(U_2=10.042\) MeV. Equilibrium deformations 0.352 and 0 yield \(\beta_{\mathrm{sc}}=0.682280\) and 0.500144. The semiaxes are (4.136108, 7.538303) and (4.932496, 7.703822) fm, and \(D=16.785589\) fm. We obtain \(V_N(0)=-16.237426\) MeV and \(V_C(0)=186.980109\) MeV. The nuclear Hessian elements \((K_{11},K_{12},K_{22})\) are (133.022448, 42.642602, 110.370319) MeV/rad², and their Coulomb counterparts are (−23.264424, −3.045934, −20.423853) MeV/rad². The full elements are (109.758024, 39.596668, 89.946467) MeV/rad². Figure 1 shows that Coulomb interaction reduces angular stiffness without destabilizing the configuration. Evaluating the mixed derivative, rather than a single diagonal stiffness, is essential for assessing the interfragment coupling.

\subsection{Inertias and the ground state of the coupled modes}\label{inertias-and-the-ground-state-of-the-coupled-modes}

Transverse inertias are approximated by those of homogeneous rigid spheroids. We use numerical units with \(\hbar=1\): inertias are in ℏ²/MeV, spins in ℏ, and energies in MeV. With \(m_Nc^2=938.91875\) MeV and \(\hbar c=197.3269804\) MeV fm,

\begin{equation}
I_i=\frac{A_i m_Nc^2(b_i^2+c_i^2)}{5(\hbar c)^2},\qquad I_0=\frac{m_Nc^2 A_1A_2D^2}{(A_1+A_2)(\hbar c)^2}.\tag{8}
\end{equation}

The relative-motion inertia \(I_0\) enters the angular-momentum constraint; no separate analysis of orbital-angular-momentum distributions is undertaken here. Eliminating the overall rotation in the sector of zero initial transverse angular momentum yields the reduced inertia matrix

\begin{equation}
M=\operatorname{diag}(I_1,I_2)-\frac{\mathbf i\mathbf i^{\mathsf T}}{I_0+I_1+I_2},\qquad \mathbf i=(I_1,I_2)^{\mathsf T}.\tag{9}
\end{equation}

The ground state of ²⁵²Cf has spin zero. For induced fission, zero total transverse angular momentum is an approximation; a more exact comparison must account for the compound-nucleus spin and its projections. Equation (9) enforces angular-momentum conservation by removing the common rotation rather than assigning independent, unconstrained fragment spins.

Let \(a_i=I_i/(I_1+I_2)\), define the wriggling coordinate \(w=a_1\theta_1+a_2\theta_2\) and the bending coordinate \(b_{\mathrm{ang}}=\theta_1-\theta_2\). Then \(\boldsymbol\theta=T(w,b_{\mathrm{ang}})^{\mathsf T}\) with

\begin{equation}
T=\begin{pmatrix}1&a_2\\1&-a_1\end{pmatrix},\qquad M_{wb}=T^{\mathsf T}MT=\operatorname{diag}(I_w,I_b),\quad K_{wb}=T^{\mathsf T}KT.\tag{10}
\end{equation}

\begin{equation}
I_w=\frac{(I_1+I_2)I_0}{I_0+I_1+I_2},\qquad I_b=\frac{I_1I_2}{I_1+I_2}.\tag{11}
\end{equation}

The bending coordinate in Ref. \cite{ref1} had a different normalization; its inertia \(I_1+(R_1/R_2)^2I_2\) must not be identified numerically with Eq. (11). The transformation here is canonical for angles and conjugate momenta. Comparisons of \(C_b\) across the two conventions require the coordinate transformation. For ⁹⁸Sr--¹³⁸Xe, \(I_1=34.942326\), \(I_2=55.690035\), \(I_0=389.333159\), \(I_w=73.518163\), and \(I_b=21.470690\) in the stated units. The transformed stiffness elements are \(\kappa_w=278.897825\), \(\kappa_b=36.049354\), and \(\kappa_{wb}=41.828652\) MeV/rad². The nonzero mixed stiffness prevents identification of the eigenmodes with two independent pure motions.

The Hamiltonian for each of the two transverse directions reads

\begin{equation}
H=\tfrac12\mathbf p^{\mathsf T}M^{-1}\mathbf p+\tfrac12\boldsymbol\theta^{\mathsf T}K\boldsymbol\theta.\tag{12}
\end{equation}

For positive-definite \(K\) and \(M\), solve the generalized eigenvalue problem

\begin{equation}
K\mathbf v_\alpha=\varepsilon_\alpha^2 M\mathbf v_\alpha,\quad \mathbf v_\alpha^{\mathsf T}M\mathbf v_\beta=\delta_{\alpha\beta},\quad \varepsilon_\alpha=\hbar\omega_\alpha.\tag{13}
\end{equation}

With \(V\) the matrix of eigenvectors and \(E=\operatorname{diag}(\varepsilon_1,\varepsilon_2)\), the ground-state momentum covariance for one transverse direction is

\begin{equation}
S=\tfrac12 MVEV^{\mathsf T}M,\qquad 4SM^{-1}S=K.\tag{14}
\end{equation}

For dimensionful momenta the right-hand side of the second identity contains \(\hbar^2\); Eq. (14) uses the numerical units specified after Eq. (8). Two equivalent transverse problems produce independent Gaussian vectors \((J_{1x},J_{2x})\) and \((J_{1y},J_{2y})\), each with covariance \(S\). The means vanish, but the mean spin magnitudes remain finite. For ⁹⁸Sr--¹³⁸Xe, the energies are 1.115446 and 2.056301 MeV; \(S_{11}=29.017153\), \(S_{12}=4.776119\), and \(S_{22}=31.979408\) in ℏ². The mean magnitudes are 6.751299 and 7.087534 ℏ.

Collective momentum covariances follow from \(S_{wb}=T^{\mathsf T}ST\). In two transverse directions, \(C_w=2(S_{wb})_{11}\), \(C_b=2(S_{wb})_{22}\), and \(C_{wb}=2(S_{wb})_{12}\). For uncoupled modes the usual relation \(C_k=I_k\hbar\omega_k\) is recovered in physical units. In the coupled problem eigenfrequencies and widths along the original collective coordinates are not interchangeable. In particular,

\begin{equation}
2S_{12}=a_1a_2C_w+(a_2-a_1)C_{wb}-C_b.\tag{15}
\end{equation}

Equation (15) shows why the ratio of eigenfrequencies alone cannot determine the sign of the correlation. Inertia fractions and mode mixing both enter; the absence of longitudinal excitation does not by itself require either sign.

\subsection{Spin distributions and the correlation sign criterion}\label{spin-distributions-and-the-correlation-sign-criterion}

The joint four-dimensional density of the spin components is

\begin{equation}
P(\mathbf J_x,\mathbf J_y)=\frac{1}{(2\pi)^2\det S}\exp\left[-\tfrac12\sum_{p=x,y}\mathbf J_p^{\mathsf T}S^{-1}\mathbf J_p\right].\tag{16}
\end{equation}

Integrating out the partner and transforming to the magnitude yields a Rayleigh distribution:

\begin{equation}
P_i(J)=\frac{J}{S_{ii}}\exp[-J^2/(2S_{ii})],\quad \langle J_i\rangle=\sqrt{\pi S_{ii}/2},\quad \langle J_i^2\rangle=2S_{ii}.\tag{17}
\end{equation}

The variance of the magnitude is \((2-\pi/2)S_{ii}\). Equation (17) is a continuous distribution of transverse angular-momentum magnitudes normalized on the positive half-line, not a discrete integer- or half-integer-spin probability \(P(J)\). The distinction is especially important at low spin and for comparisons with exact angular-momentum projection. We compare arithmetic means rather than effective spins defined from \(J(J+1)\).

Let \(\rho=S_{12}/\sqrt{S_{11}S_{22}}\) be the correlation of like Cartesian components. In Ref. \cite{ref1} the Pearson coefficient refers to spin magnitudes. For normalized magnitudes \(r=J_1/\sqrt{S_{11}}\), \(t=J_2/\sqrt{S_{22}}\), Eq. (16) gives

\begin{equation}
p(r,t)=\frac{rt}{1-\rho^2}\exp\left[-\frac{r^2+t^2}{2(1-\rho^2)}\right] I_0^{(B)}\left(\frac{\rho rt}{1-\rho^2}\right),\tag{18}
\end{equation}

Here \(I_0^{(B)}\) is the modified Bessel function (not the relative-motion inertia). Expanding it in a series and integrating the terms over \(r,t\) using gamma functions, followed by a hypergeometric transformation, gives

\begin{equation}
\langle J_1J_2\rangle=\frac{\pi}{2}\sqrt{S_{11}S_{22}}\,F_{2,1}(-\tfrac12,-\tfrac12;1;\rho^2).\tag{19}
\end{equation}

Here \(F_{2,1}\) denotes the Gauss hypergeometric function, conventionally written ₂F₁. The within-channel Pearson coefficient of the magnitudes is then

\begin{equation}
c(\rho)=\frac{\pi}{4-\pi}\left[F_{2,1}(-\tfrac12,-\tfrac12;1;\rho^2)-1\right].\tag{20}
\end{equation}

For small \(\rho\), \(c\simeq\pi\rho^2/[4(4-\pi)]\). Both \(\rho=+0.16\) and \(\rho=-0.16\) yield \(c\simeq0.02346\). Equation (20) is even and nonnegative for a nondegenerate planar Gaussian pair; this is not a property of arbitrary spin statistics. Related mathematical distributions have been examined in other fields \cite{ref20}. Here they provide the normalized map from a specified angular Hamiltonian to spin observables.

Let the opening angle be \(\varphi\in[0,\pi]\). Integrating over the common azimuth and including both branches of the oriented angle gives

\begin{equation}
p(r,t,\varphi)=\frac{rt}{\pi(1-\rho^2)}\exp\left[-\frac{r^2+t^2-2\rho rt\cos\varphi}{2(1-\rho^2)}\right].\tag{21}
\end{equation}

To integrate over the magnitudes, substitute \(r=\sqrt{1-\rho^2}u\), \(t=\sqrt{1-\rho^2}v\), and \(z=\rho\cos\varphi\). The integral of \(uv\exp[-(u^2+v^2-2zuv)/2]\) is the derivative with respect to \(z\) of the quadrant Gaussian integral

\begin{equation}
\mathcal I(z)=\int_0^\infty\int_0^\infty e^{-(u^2+v^2-2zuv)/2}\,du\,dv=\frac{\pi/2+\arcsin z}{\sqrt{1-z^2}}.\tag{22}
\end{equation}

The opening-angle density per radian is therefore

\begin{equation}
P(\varphi)=\frac{1-\rho^2}{\pi}\left[\frac{1}{1-z^2}+\frac{z(\pi/2+\arcsin z)}{(1-z^2)^{3/2}}\right].\tag{23}
\end{equation}

For \(\rho=0\) the density is uniform. To evaluate the obtuse-angle probability, write the second normalized vector as \(Y=\rho X+\sqrt{1-\rho^2}Z\), where \(X,Z\) are independent standard planar Gaussian vectors. Conditional on \(r=(X^{\mathsf T}X)^{1/2}\), the probability of a negative scalar product is \(\Phi[-\rho r/\sqrt{1-\rho^2}]\), with \(\Phi\) the standard normal cumulative distribution. Integration over the Rayleigh distribution and integration by parts yield

\begin{equation}
P_{\mathrm{obt}}=\Pr(\varphi>\pi/2)=\frac{1-\rho}{2}.\tag{24}
\end{equation}

Equations (20) and (24) establish a qualitative difference between the observables: the magnitude correlation is quadratic at small \(\rho\), whereas the acute--obtuse asymmetry is linear. At \(\rho=0.16\) the obtuse fraction is 42\%, and at \(\rho=-0.16\) it is 58\%, despite identical magnitude correlations. The computational supplement includes double quadrature checks of Eq. (23) and Monte Carlo checks of Eqs. (20) and (24), using 500,000 events for each of four values of \(\rho\). At 12 control points, the maximum quadrature discrepancy from the analytic density is below \(10^{-8}\); Monte Carlo deviations are compatible with finite sampling.

To infer the sign directly from the Hamiltonian, use the square root of a positive-definite two-dimensional matrix. Let \(A=M^{-1/2}KM^{-1/2}\) and \(d=\sqrt{\det A}=\varepsilon_1\varepsilon_2\). The Cayley--Hamilton identity gives

\[A^{1/2}=\frac{A+d\,\mathbb{1}}{\sqrt{\operatorname{tr}A+2d}}.\]

Substitution into \(S=\tfrac12 M^{1/2}A^{1/2}M^{1/2}\) gives a closed form independent of numerical diagonalization:

\begin{equation}
S=\frac{K+dM}{2\sqrt{\operatorname{tr}(M^{-1}K)+2d}}.\tag{25}
\end{equation}

By Eq. (25) the off-diagonal covariance is governed by competing terms \(K_{12}\) and \(dM_{12}\), with \(M_{12}<0\) for the reduced inertia. Define

\begin{equation}
k=\frac{K_{12}}{\sqrt{K_{11}K_{22}}},\qquad m=-\frac{M_{12}}{\sqrt{M_{11}M_{22}}}.\tag{26}
\end{equation}

Positive definiteness requires \(-1<k<1\) and \(0<m<1\). Then

\begin{equation}
\frac{K_{12}+dM_{12}}{\sqrt{K_{11}K_{22}}}=k-m\sqrt{\frac{1-k^2}{1-m^2}}.\tag{27}
\end{equation}

For \(k<0\), Eq. (27) is negative. For \(k\geq0\), comparing the squares of the two nonnegative terms establishes that the sign of Eq. (27) equals that of \(k-m\). Hence the following condition is exact within the coupled harmonic problem:

\begin{equation}
\operatorname{sgn}\rho=\operatorname{sgn}(k-m),\qquad P_{\mathrm{obt}}=\frac{1-\rho}{2}\tag{28}
\end{equation}

For \(k>m\), the obtuse fraction is below one half; at equality it is one half; for \(k<m\) it exceeds one half. Kinematic and potential coupling compete not only in the mode energies but also in the sign of the spatial correlation. This criterion does not require fitting another model's angular distributions. It assumes positive-definite matrices, equivalent transverse directions, zero means, and the harmonic ground state; it is not a general theorem about three-dimensional spins, thermal mixtures, non-Gaussian densities, or exact quantum projection.

\section{Inputs and choice of prescission configurations}\label{inputs-and-choice-of-prescission-configurations}

\subsection{Primary channels and weights}\label{primary-channels-and-weights}

The archived tables for ²³⁶U and ²⁴⁰Pu contain 191 and 206 complementary pairs, respectively: 397 pairs and 793 individual AMPLIFI distributions (the symmetric pair has only one unique fragment type). The direct calculation takes masses, charges, neutron multiplicities, and separations from the previously prepared inputs. Complete inputs are available for 327 pairs; we do not invent geometries for the remaining 70. AMPLIFI spins are not target values in the evaluation of the angular Hessian. Nevertheless, the provenance of every archived separation is not fully independent of earlier work. The calculation is direct conditional on its inputs, rather than a parameter-free prediction of the entire fission process.

Positive archived weights are available for 62 and 71 uranium and plutonium pairs. A zero archived weight denotes the absence of a positive entry in the adopted table, not a demonstrated prohibition of a physical channel. All 327 solutions are retained for pairwise comparisons; only positively weighted channels enter the integrated distributions. AMPLIFI comparisons use the same normalized weights on the common support and arithmetic means of the individual spins.

For the three Wilson systems \cite{ref2}, use the published even-even residual isotopes and their individual yields in Extended Data Tables 1--3 (29, 33, and 31 isotopes for ²³³Th, ²³⁹U, and ²⁵²Cf). Complementary charges and mass balance, assuming no pre-fission emission, define candidate channels. For residual masses \(A_L^r,A_H^r\), the total number of neutrons is \(N=A_{\mathrm{CN}}-A_L^r-A_H^r\). Consider \(0\leq N\leq8\) and every integer allocation \(n_L+n_H=N\). The primary masses are \(A_i=A_i^r+n_i\). Deformations, inertias, and stiffnesses are evaluated for each allocation without using the measured spin in Eqs. (1)--(14).

In the absence of published joint coincidence yields, adopt the conditional model weight

\begin{equation}
w_q^{\mathrm{raw}}=Y_L^rY_H^r\binom{N}{n_L}2^{-N},\qquad w_q=\frac{w_q^{\mathrm{raw}}}{\sum_{q\in\mathcal S}w_q^{\mathrm{raw}}}.\tag{29}
\end{equation}

Here \(\mathcal S\) is the support of the calculation. Products of individual yields and binomial neutron partitions are not measured joint yields. They provide a reproducible scenario for investigating the Hamiltonian but do not reconstruct the actual correlations of mass, charge, energy, and multiplicity. Since Eq. (29) uses only observed isotopes, its integrated results are not unconditional physical-yield averages. The weights must be checked against primary coincidence data. Residual isotope averages use conditional normalization over the channels feeding the isotope in question.

\begin{table}[htbp]
\centering
\caption{Counts of input records, stable baseline solutions, and positively weighted solutions. In the Wilson systems the channels are possible neutron allocations, not distinct measured pairs.}
\small
\begin{adjustbox}{max width=\linewidth}
\begin{tabular}{lccc}
\toprule
System & Inputs & Stable solutions & Positive weights \\
\midrule
²³³Th & 154 & 154 & 154 \\
²³⁶U & 191 & 167 & 62 \\
²³⁹U & 208 & 208 & 208 \\
²⁴⁰Pu & 206 & 160 & 71 \\
²⁵²Cf & 150 & 150 & 150 \\
Total & 909 & 839 & 645 \\
\bottomrule
\end{tabular}\end{adjustbox}
\end{table}

\subsection{Geometry and energy constraints}\label{geometry-and-energy-constraints}

The baseline geometry for the Wilson channels uses a 1.5-fm gap between the facing longitudinal surfaces; archived separations are used for ²³⁶U and ²⁴⁰Pu. The direct calculation shows that Coulomb energy alone exceeds the available energy for some conditional channels. Rather than alter shapes or frequencies arbitrarily, we impose an additional condition on the configuration energy.

The reaction energy \(Q\) is calculated from AME2020 masses \cite{ref10}, including the entrance energy. For neutron-induced Wilson systems, a mean laboratory neutron energy of 1.9 MeV is transformed to the center-of-mass frame; no incident energy is added for spontaneous fission. Assuming zero initial radial kinetic energy and omitting a separate zero-point contribution in the energy balance, solve

\begin{equation}
V_N(D,0)+V_C(D,0)=Q-U_L-U_H.\tag{30}
\end{equation}

The two radial branches are treated separately. A root must also have positive-definite \(K\) to be angularly stable. The inner branch yields 452 refined solutions, whereas the outer branch is predominantly angularly unstable. The maximum absolute residual of the refined energy equality is below 0.0012 MeV. This is the numerical accuracy of the assumed equation, not the completeness of the physical energy balance.

Of the 154, 208, and 150 baseline channels, 145, 189, and 118 remain on the inner branch, respectively. Their retained model-weight fractions are 96.78\%, 73.35\%, and 65.63\%. The other channels are excluded from inner-branch averages. To separate geometric effects from selection effects, the baseline and refined solutions are also compared on matching supports with identical weights. The 1291 solutions (839 baseline plus 452 alternatives) are not 1291 distinct physical channels.

\section{Results}\label{results}

\subsection{Inertias, stiffnesses, and energies of bending and wriggling motion}\label{inertias-stiffnesses-and-energies-of-bending-and-wriggling-motion}

Table II gives weighted means for the five systems. Equation (13) is solved for each channel before averaging. Substituting averaged matrices into the nonlinear Eq. (25) does not generally recover the mean covariance; the same consideration applies to mean spins and correlation coefficients.

\begin{table}[htbp]
\centering
\caption{Mean inertias (ℏ²/MeV), angular stiffnesses (MeV/rad²), and collective momentum widths (ℏ²). Indices \(w,b\) refer to the coordinates of Eq. (10); \(wb\) denotes the mixed element.}
\small
\begin{adjustbox}{max width=\linewidth}
\begin{tabular}{lcccccccc}
\toprule
System & \(I_w\) & \(I_b\) & \(\kappa_w\) & \(\kappa_b\) & \(\kappa_{wb}\) & \(C_w\) & \(C_b\) & \(C_{wb}\) \\
\midrule
²³³Th & 71.87 & 19.63 & 277.44 & 36.45 & 41.72 & 138.21 & 25.73 & 12.82 \\
²³⁶U & 74.03 & 20.15 & 266.77 & 34.35 & 39.88 & 138.31 & 25.44 & 12.72 \\
²³⁹U & 78.04 & 22.23 & 321.82 & 38.15 & 43.04 & 155.74 & 28.13 & 13.19 \\
²⁴⁰Pu & 77.85 & 22.58 & 334.83 & 38.76 & 41.04 & 159.12 & 28.65 & 12.18 \\
²⁵²Cf & 85.98 & 24.98 & 337.46 & 36.92 & 36.24 & 168.08 & 29.59 & 11.50 \\
\bottomrule
\end{tabular}\end{adjustbox}
\end{table}

The mean inertias increase with system mass, but this alone does not fix the energy changes. For ²⁴⁰Pu, both mean bending and mixed stiffnesses are larger than for ²³⁶U. The mixed stiffness of ²⁵²Cf is smaller than for the other systems despite larger inertias. Neither mass nor deformation alone is a universal parameter for spin orientation.

To characterize the mode composition, calculate eigenvectors in the mass-normalized coordinates \(M_{wb}^{1/2}(w,b_{\mathrm{ang}})^{\mathsf T}\). Define \(f_b\) as the squared bending component of the normalized lower-mode eigenvector. Removing the mixed stiffness yields the pure energies \(\varepsilon_b^{(0)}=\sqrt{\kappa_b/I_b}\) and \(\varepsilon_w^{(0)}=\sqrt{\kappa_w/I_w}\). Comparison with the coupled eigenenergies isolates the effect of mixing at fixed inertias and diagonal stiffnesses.

\begin{table}[htbp]
\centering
\caption{Mean coupled eigenenergies and pure-mode energies without the mixed stiffness (MeV); \(f_b\) is the mean bending fraction of the lower mode.}
\small
\begin{adjustbox}{max width=\linewidth}
\begin{tabular}{lccccc}
\toprule
System & \(\varepsilon_1\) & \(\varepsilon_2\) & \(\varepsilon_b^{(0)}\) & \(\varepsilon_w^{(0)}\) & \(f_b\), \% \\
\midrule
²³³Th & 1.140 & 2.077 & 1.356 & 1.943 & 83.1 \\
²³⁶U & 1.119 & 2.014 & 1.307 & 1.897 & 83.6 \\
²³⁹U & 1.132 & 2.113 & 1.307 & 2.009 & 86.9 \\
²⁴⁰Pu & 1.170 & 2.142 & 1.309 & 2.059 & 89.5 \\
²⁵²Cf & 1.097 & 2.024 & 1.212 & 1.956 & 90.8 \\
\bottomrule
\end{tabular}\end{adjustbox}
\end{table}

\begin{figure}[htbp]
\centering
\includegraphics[width=0.98\linewidth,height=.65\textheight,keepaspectratio]{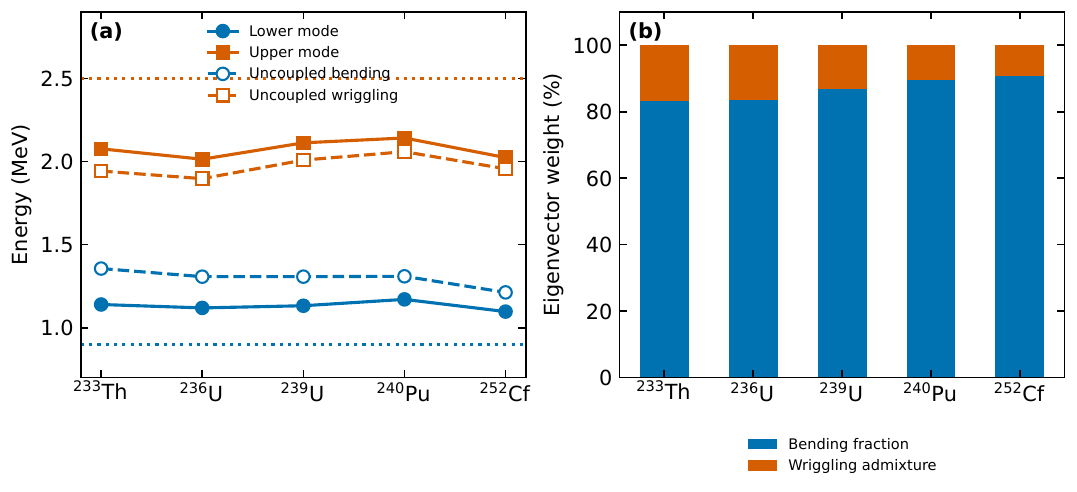}
\caption{(a) Coupled and uncoupled energies; dotted horizontal lines denote the fixed energies 0.9 and 2.5 MeV used in Ref. \cite{ref1}. (b) Bending and wriggling components of the mass-normalized lower-mode eigenvector. Plotting code was prepared with assistance from ChatGPT (OpenAI); curves and markers are based on the stated numerical inputs.}
\end{figure}

Mixing increases the level splitting relative to the uncoupled modes. In the baseline geometry, the lower energy exceeds the former fixed 0.9 MeV, whereas the higher energy is below 2.5 MeV. The lower mode is 83--91\% bending, permitting the bending/wriggling terminology provided that the admixture is specified. Eigenenergies are invariant under linear coordinate renormalization; numerical widths \(C_b,C_w\) need not be. Comparing energies with Ref. \cite{ref1} is thus less ambiguous than comparing individual inertias defined in different coordinate conventions.

\subsection{Sign of the spin-vector correlation}\label{sign-of-the-spin-vector-correlation}

Equation (25) is verified for all 1291 stable solutions. The maximum absolute elementwise discrepancy from independently diagonalized covariances is \(8.53\times10^{-14}\) ℏ². All solutions have \(k>m\), including those without positive archived weights. The minimum margin \(k-m\) is 0.13495. The calculated sign follows directly from the input Hamiltonian without fitting angular distributions of other theories. This conclusion applies to the examined stable configurations, not to all conceivable cold scission configurations.

\begin{table}[htbp]
\centering
\caption{Mean normalized matrix elements and obtuse-angle fractions in the coupled problem, after removing mixing in \((w,b_{\mathrm{ang}})\) coordinates, and with fixed energies \cite{ref1} but the new inertias. The final column is a diagnostic control, not a reproduction of the former Migdal model.}
\small
\begin{adjustbox}{max width=\linewidth}
\begin{tabular}{lcccccc}
\toprule
System & \(\bar k\) & \(\bar m\) & \(\overline{k-m}\) & Coupled, \% & Uncoupled, \% & 0.9/2.5 MeV, \% \\
\midrule
²³³Th & 0.368 & 0.103 & 0.265 & 43.03 & 46.84 & 32.05 \\
²³⁶U & 0.391 & 0.101 & 0.290 & 42.34 & 46.34 & 32.00 \\
²³⁹U & 0.402 & 0.101 & 0.301 & 42.01 & 44.86 & 31.44 \\
²⁴⁰Pu & 0.414 & 0.101 & 0.313 & 41.67 & 43.99 & 31.22 \\
²⁵²Cf & 0.414 & 0.101 & 0.313 & 41.63 & 43.59 & 31.20 \\
\bottomrule
\end{tabular}\end{adjustbox}
\end{table}

\begin{figure}[htbp]
\centering
\includegraphics[width=0.98\linewidth,height=.65\textheight,keepaspectratio]{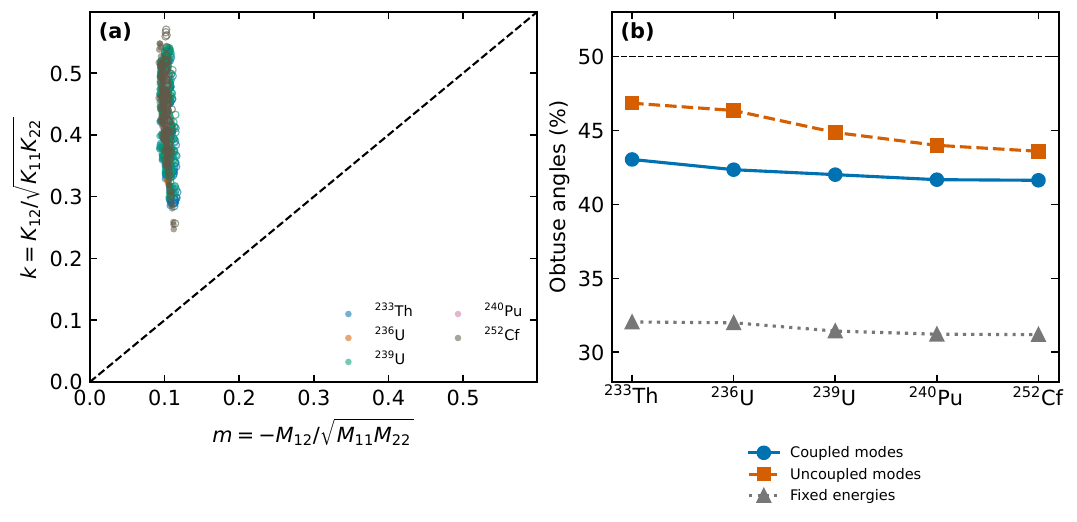}
\caption{(a) Verification of Eq. (28) for baseline (filled) and refined (open) configurations. The diagonal marks equal acute and obtuse fractions. (b) Full coupled calculation, artificially uncoupled modes, and fixed energies with the new inertias. Plotting code was prepared with assistance from ChatGPT (OpenAI); curves and markers are based on the stated numerical inputs.}
\end{figure}

Neither returning to fixed energies nor removing the mixed stiffness at fixed diagonal stiffnesses produces a predominance of obtuse angles in the weighted mixtures. The fixed energies instead strengthen the acute-angle preference. This comparison separates the role of frequencies and mixing from the changes in the inertia prescription and collective-coordinate conventions. It does not imply that the curves published in Ref. \cite{ref1} should coincide with the fixed-energy control. Changing the bending-to-wriggling width ratio can reverse the sign in Eq. (15) without a twisting mode; Eq. (28) specifies the corresponding condition on the full Hamiltonian.

\subsection{Primary spins and magnitude correlations}\label{primary-spins-and-magnitude-correlations}

For each configuration use Eq. (17); the mixture is \(P_i(J)=\sum_qw_qP_{i,q}(J)\). The first and second moments are \(\sum_qw_q\sqrt{\pi S_{ii,q}/2}\) and \(2\sum_qw_qS_{ii,q}\), respectively. A mixture width need not equal a single-channel width. Figure 4 shows the five systems, including the high-spin tail. Inner-branch results are included for the three Wilson systems. Refined geometry raises the mode energies and spin magnitudes, but has a much weaker effect on the angular correlation.

\begin{figure}[htbp]
\centering
\includegraphics[width=0.98\linewidth,height=.65\textheight,keepaspectratio]{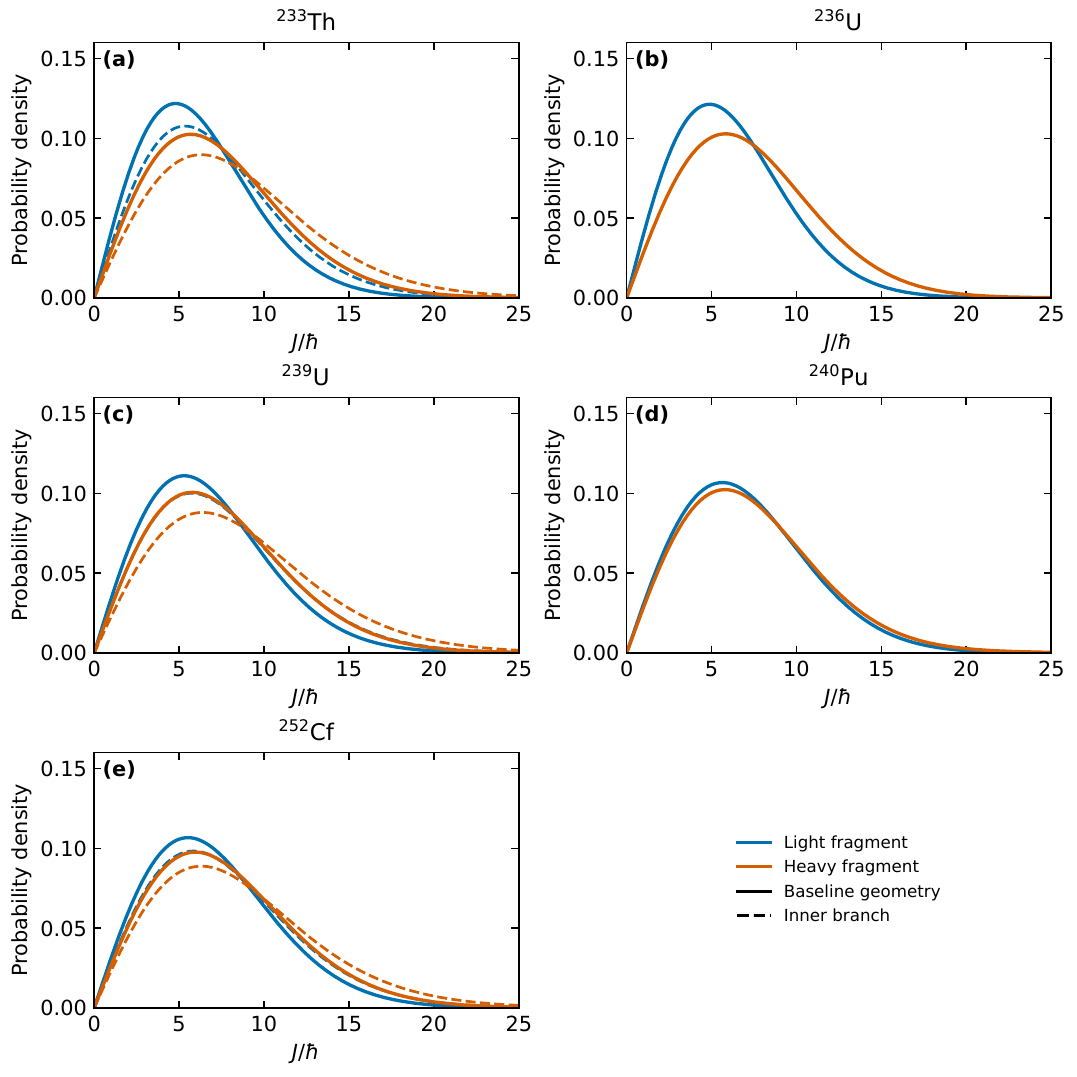}
\caption{Primary light- and heavy-fragment spin-magnitude distributions. Solid lines: baseline; dashed: stable inner branch for the Wilson systems. The distributions are continuous densities, not discrete projected spin probabilities. Plotting code was prepared with assistance from ChatGPT (OpenAI); curves and markers are based on the stated numerical inputs.}
\end{figure}

Ref. \cite{ref1} averaged the pairwise correlations with yield weights. For that comparison, define \(\bar c=\sum_qw_qc_q\). For an entire event mixture, the mixed moments must instead be calculated before forming the Pearson coefficient. With \(\mu_{L,q},\mu_{H,q}\) the conditional mean magnitudes, the law of total covariance gives

\begin{equation}
\operatorname{cov}_{\mathrm{mix}}(J_L,J_H)=\sum_q w_q\operatorname{cov}_q(J_L,J_H)+\operatorname{cov}_w(\mu_{L,q},\mu_{H,q}).\tag{31}
\end{equation}

\begin{equation}
\operatorname{var}_{\mathrm{mix}}(J_i)=\sum_q w_q\operatorname{var}_q(J_i)+\operatorname{var}_w(\mu_{i,q}).\tag{32}
\end{equation}

The mixture coefficient \(c_{\mathrm{mix}}\) follows from these covariance and variance terms and need not equal \(\bar c\). The first term of Eq. (31) is nonnegative for each Gaussian pair; the second depends on the change of conditional means across mass-charge channels. Anticorrelated conditional means can compensate positive within-channel covariance.

\begin{table}[htbp]
\centering
\caption{Mean primary spins (ℏ), component correlation, mean pairwise magnitude correlation, and magnitude correlation of the full mixture for positively weighted baseline solutions.}
\small
\begin{adjustbox}{max width=\linewidth}
\begin{tabular}{lccccc}
\toprule
System & \(\langle J_L\rangle\) & \(\langle J_H\rangle\) & \(\bar\rho\) & \(\bar c\) & \(c_{\mathrm{mix}}\) \\
\midrule
²³³Th & 6.238 & 7.410 & 0.1393 & 0.0191 & 0.0032 \\
²³⁶U & 6.264 & 7.393 & 0.1531 & 0.0215 & −0.0006 \\
²³⁹U & 6.844 & 7.549 & 0.1598 & 0.0248 & 0.0125 \\
²⁴⁰Pu & 7.118 & 7.455 & 0.1667 & 0.0256 & 0.0095 \\
²⁵²Cf & 7.123 & 7.780 & 0.1675 & 0.0273 & 0.0217 \\
\bottomrule
\end{tabular}\end{adjustbox}
\end{table}

\begin{figure}[htbp]
\centering
\includegraphics[width=0.98\linewidth,height=.65\textheight,keepaspectratio]{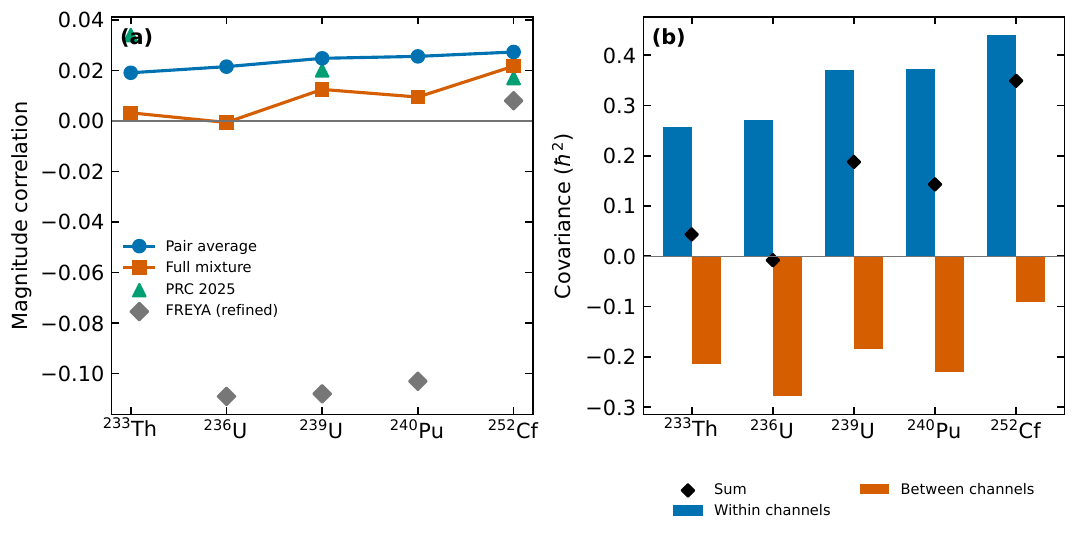}
\caption{(a) Average channel magnitude correlations and full-mixture coefficients, with values from Refs. \cite{ref1,ref3}. (b) Within-channel and between-channel contributions to the total covariance in Eq. (31). The distinct averaging definitions are explicitly distinguished. Plotting code was prepared with assistance from ChatGPT (OpenAI); curves and markers are based on the stated numerical inputs.}
\end{figure}

Ref. \cite{ref1} gave 0.034, 0.020, and 0.017 for ²³³Th, ²³⁹U, and ²⁵²Cf. The new \(\bar c\) values are 0.0191, 0.0248, and 0.0273. Thus changes in the angular dynamics do not produce large magnitude correlations, although the numerical values change. This comparison is not event matched: samples, geometries, and weights differ.

The ²³⁶U case illustrates the mixture effect particularly clearly. With mean channel correlation 0.0215, the within-channel covariance is +0.27085 ℏ², whereas the between-channel contribution is −0.27861 ℏ². Their near cancellation gives \(c_{\mathrm{mix}}=-0.00059\). The corresponding pairs of contributions are (0.2580, −0.2145), (0.3716, −0.1839), (0.3736, −0.2304), and (0.4401, −0.0910) ℏ² for ²³³Th, ²³⁹U, ²⁴⁰Pu, and ²⁵²Cf. A small mixture coefficient therefore does not require small within-channel covariance. This is a statistical mixture effect, not an additional spin-generation mechanism.

In the original FREYA calculation \cite{ref3}, full-population magnitude correlations are 0.002, 0.002, 0.001, and 0.001 for ²³⁶U, ²³⁹U, ²⁴⁰Pu, and ²⁵²Cf; with refined inertias they are −0.109, −0.108, −0.103, and 0.008. These must be compared with \(c_{\mathrm{mix}}\), not \(\bar c\). The refined uranium and plutonium FREYA values differ appreciably from our mixtures. A single similar small coefficient does not make the models equivalent. For the Wilson systems, Eq. (29) directly affects the between-channel covariance and limits quantitative comparisons.

\subsection{Spatial orientation of the spin vectors}\label{spatial-orientation-of-the-spin-vectors}

The mixture angle density is \(\sum_qw_qP_q(\varphi)\). By linearity and Eq. (24), \(\bar P_{\mathrm{obt}}=(1-\bar\rho)/2\) exactly. The full density cannot be recovered from \(\bar\rho\) alone, because Eq. (23) is nonlinear. Figure 6 uses a mixture of channel densities, not the density at the average coefficient. The five numerical normalizations differ from unity by less than \(10^{-14}\).

\begin{figure}[htbp]
\centering
\includegraphics[width=0.98\linewidth,height=.65\textheight,keepaspectratio]{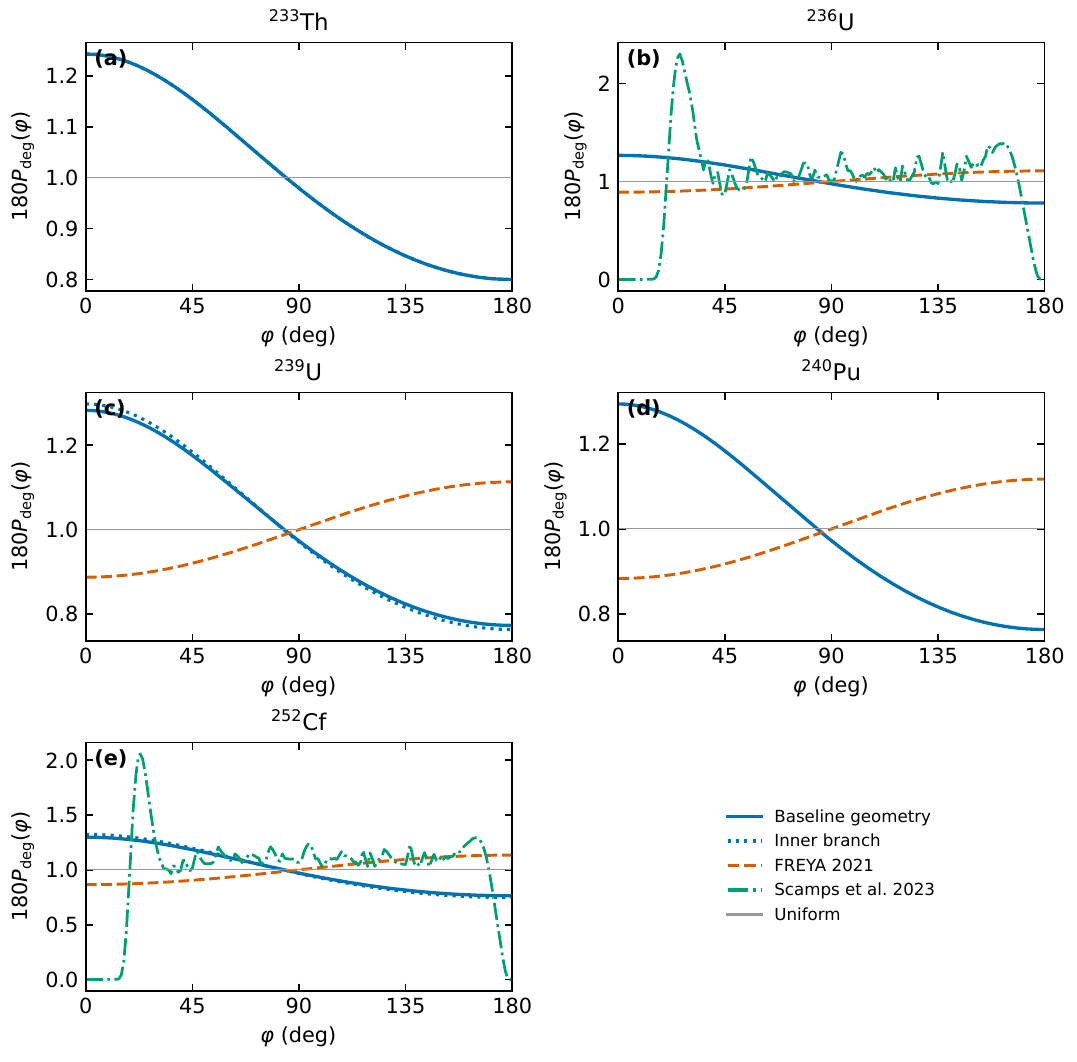}
\caption{Weighted angle densities for five systems. The baseline and inner-branch calculations are compared with FREYA \cite{ref3} and digitized continuous curves of Scamps et al.~\cite{ref6}. The ordinate is 180 times the density per degree; unity represents uniform planar orientation. Plotting code was prepared with assistance from ChatGPT (OpenAI); curves and markers are based on the stated numerical inputs.}
\end{figure}

\begin{table}[htbp]
\centering
\caption{Obtuse-angle fractions. FREYA values use the first harmonic in the refined parameterization \cite{ref3}; Scamps values integrate the normalized smooth curves of Fig. 3 of Ref. \cite{ref6}, excluding the discrete contribution at 180°. A dash indicates no comparable result in the examined sources.}
\small
\begin{adjustbox}{max width=\linewidth}
\begin{tabular}{lccc}
\toprule
System & Present, \% & FREYA, \% & Scamps, \% \\
\midrule
²³³Th & 43.03 & --- & --- \\
²³⁶U & 42.34 & 53.47 & 52.06 \\
²³⁹U & 42.01 & 53.60 & --- \\
²⁴⁰Pu & 41.67 & 53.72 & --- \\
²⁵²Cf & 41.63 & 54.30 & 52.23 \\
\bottomrule
\end{tabular}\end{adjustbox}
\end{table}

For FREYA we use \(P(\varphi)=(1+f_1\cos\varphi)/\pi\) with \(f_1=-0.109,-0.113,-0.117,-0.135\) \cite{ref3}. Integration over \(\pi/2<\varphi\leq\pi\) gives \(1/2-f_1/\pi\). The asymmetry differs from ours in both magnitude and sign. A similar difference is found with Scamps et al.~The total-variation distances between normalized continuous curves for ²³⁶U and ²⁵²Cf are 18.1\% and 16.7\%; this is a descriptive distance, not a statistical significance level.

The evolution of microscopic estimates matters. The earlier calculation of Bulgac et al.~\cite{ref5} reported approximately 71--72\% obtuse angles for ²⁵²Cf; the more complete joint projection in Ref. \cite{ref6} yields about 53\%. Table VI uses smooth curves, so the authors' approximately 53\% is preferable for the full result. The revised projection reduces but does not eliminate the difference from our model. We use continuous transverse vectors; Ref. \cite{ref6} constructs the angle from coupled angular-momentum quantum numbers and permits longitudinal projections. This comparison diagnoses differences of models and observables but cannot replace a common quantum projection.

Ref. \cite{ref1} also considered the relative difference of the densities at 180° and 0°. In the new baseline mixtures, \([P(\pi)-P(0)]/[P(\pi)+P(0)]\) equals −21.7\%, −24.8\%, and −26.0\% for ²³³Th, ²³⁹U, and ²⁵²Cf. The former publication reported 13.5\%, about 2\%, and 12\%, with different asymmetry directions. The endpoint contrast is not the obtuse-angle fraction; the accessible preprint does not unambiguously specify the sign of the second numerical value, so we do not assign one. Weak magnitude correlations persist, whereas the orientation preference changes, notably for ²³³Th. In Eqs. (7)--(9) of that preprint, the last line of Eq. (8) has an inconsistent written sign for the bending contribution. Here Eqs. (10) and (15) fix the sign canonically. The typographical inconsistency alone does not establish an error in the earlier numerical implementation.

\subsection{Primary AMPLIFI spins and residual Wilson spins}\label{primary-amplifi-spins-and-residual-wilson-spins}

The comparison with AMPLIFI uses the arithmetic mean \(\sum_J JP(J)\) for the primary isotope. It is not the effective spin defined by \(J_{\mathrm{eff}}(J_{\mathrm{eff}}+1)=\sum_J J(J+1)P(J)\). Our continuous magnitude in Eq. (17) is not identical to the discrete quantum number, but the comparison provides a scale of individual discrepancies without fitting the microscopic spins.

\begin{table}[htbp]
\centering
\caption{Mean primary spins on common positively weighted support (ℏ). RMS uses both partners with the adopted weights; the last column counts pairs whose two relative deviations are at most 10\%.}
\small
\begin{adjustbox}{max width=\linewidth}
\begin{tabular}{lccccccc}
\toprule
System & Pairs & Present light & AMPLIFI light & Present heavy & AMPLIFI heavy & RMS & Both ≤10\% \\
\midrule
²³⁶U & 62 & 6.26 & 6.26 & 7.39 & 6.22 & 1.815 & 4 \\
²⁴⁰Pu & 71 & 7.12 & 7.46 & 7.46 & 4.48 & 2.624 & 4 \\
\bottomrule
\end{tabular}\end{adjustbox}
\end{table}

\begin{figure}[htbp]
\centering
\includegraphics[width=0.98\linewidth,height=.65\textheight,keepaspectratio]{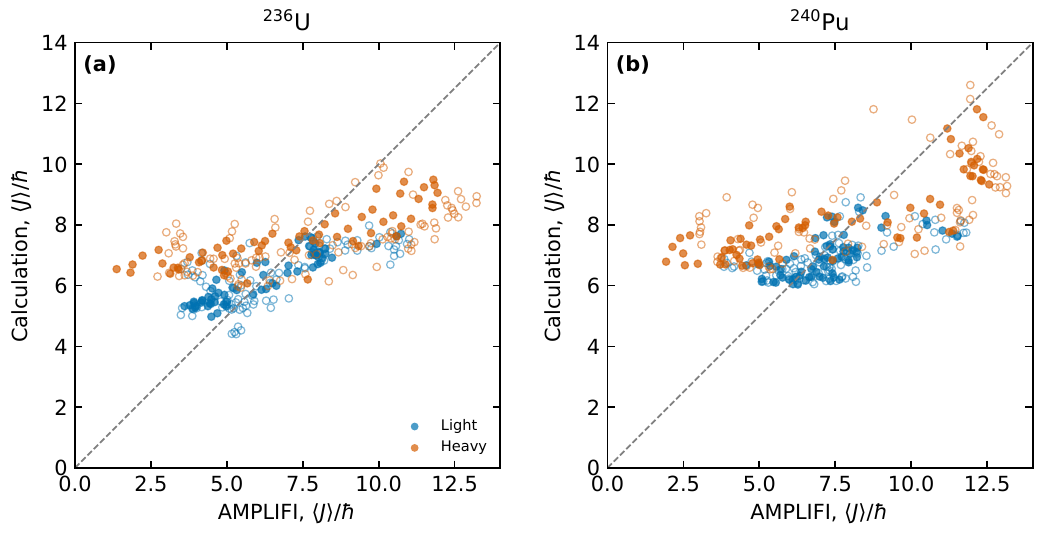}
\caption{Pairwise comparison of primary spins with AMPLIFI. Filled markers denote positively weighted pairs; open markers denote the other available solutions. The dashed line marks equality. Plotting code was prepared with assistance from ChatGPT (OpenAI); curves and markers are based on the stated numerical inputs.}
\end{figure}

For all 167 and 160 available pairs with equal weights, the RMS deviations are 2.098 and 2.164 ℏ; 7 and 6 pairs satisfy the 10\% criterion for both spins. Close average light-fragment spins do not imply reproduction of the individual distributions. The heavy fragments of ²⁴⁰Pu are especially discrepant. Rigid inertias, incomplete shell structure, and deformation energy may contribute, but the present comparison does not separate them. Refined FREYA light/heavy means \cite{ref3} are 6.08/5.31, 6.59/5.49, 6.86/5.60, and 7.48/6.80 ℏ for ²³⁶U, ²³⁹U, ²⁴⁰Pu, and ²⁵²Cf. These refer to different populations but also motivate scrutiny of the heavy-fragment spin scale.

Wilson spins \cite{ref2} refer to the stage after neutron evaporation with an estimated statistical-photon correction included in the extraction. Subtracting that photon decrement again would double count it. For conditional primary channel \(q\), use a simple mean residual-spin prescription

\begin{equation}
J_{i,q}^{r}(\alpha)=\langle J_{i,q}\rangle-\alpha n_{i,q}.\tag{33}
\end{equation}

Here \(\alpha\) is the effective signed decrement of mean spin magnitude per neutron, not the mean magnitude of the neutron angular momentum. Individual evaporation events may increase the daughter-spin magnitude \cite{ref11}. Equation (33) is a mean law and may yield unphysical values in individual channels for large decrements. Table VIII preserves the linear, rather than zero-truncated, prescription of the original calculations; it does not replace an event cascade. We test three values used in previous group studies, 0.536, 1.10, and 1.23 ℏ/neutron, without refitting. They come from different calibrations and are sensitivity scenarios rather than a single consistent estimate. The \(\alpha=0\) control omits neutron attenuation.

\begin{table}[htbp]
\centering
\caption{RMS deviations of calculated residual spins from Wilson values (ℏ) for different decrements. Parentheses give isotope counts; a dash denotes an unevaluated scenario.}
\small
\begin{adjustbox}{max width=\linewidth}
\begin{tabular}{lcccc}
\toprule
System and geometry & \(\alpha=0\) & 0.536 & 1.10 & 1.23 \\
\midrule
²³³Th, baseline (29) & 2.129 & 1.508 & 1.136 & 1.125 \\
²³⁹U, baseline (32) & 1.957 & 1.279 & 1.244 & 1.361 \\
²⁵²Cf, baseline (29) & 1.484 & 1.220 & 1.609 & 1.764 \\
²³³Th, inner (29) & --- & 2.205 & 1.462 & 1.321 \\
²³⁹U, inner (32) & --- & 1.912 & 1.143 & 1.080 \\
²⁵²Cf, inner (27) & --- & 1.460 & 1.252 & 1.345 \\
\bottomrule
\end{tabular}\end{adjustbox}
\end{table}

\begin{figure}[htbp]
\centering
\includegraphics[width=0.98\linewidth,height=.65\textheight,keepaspectratio]{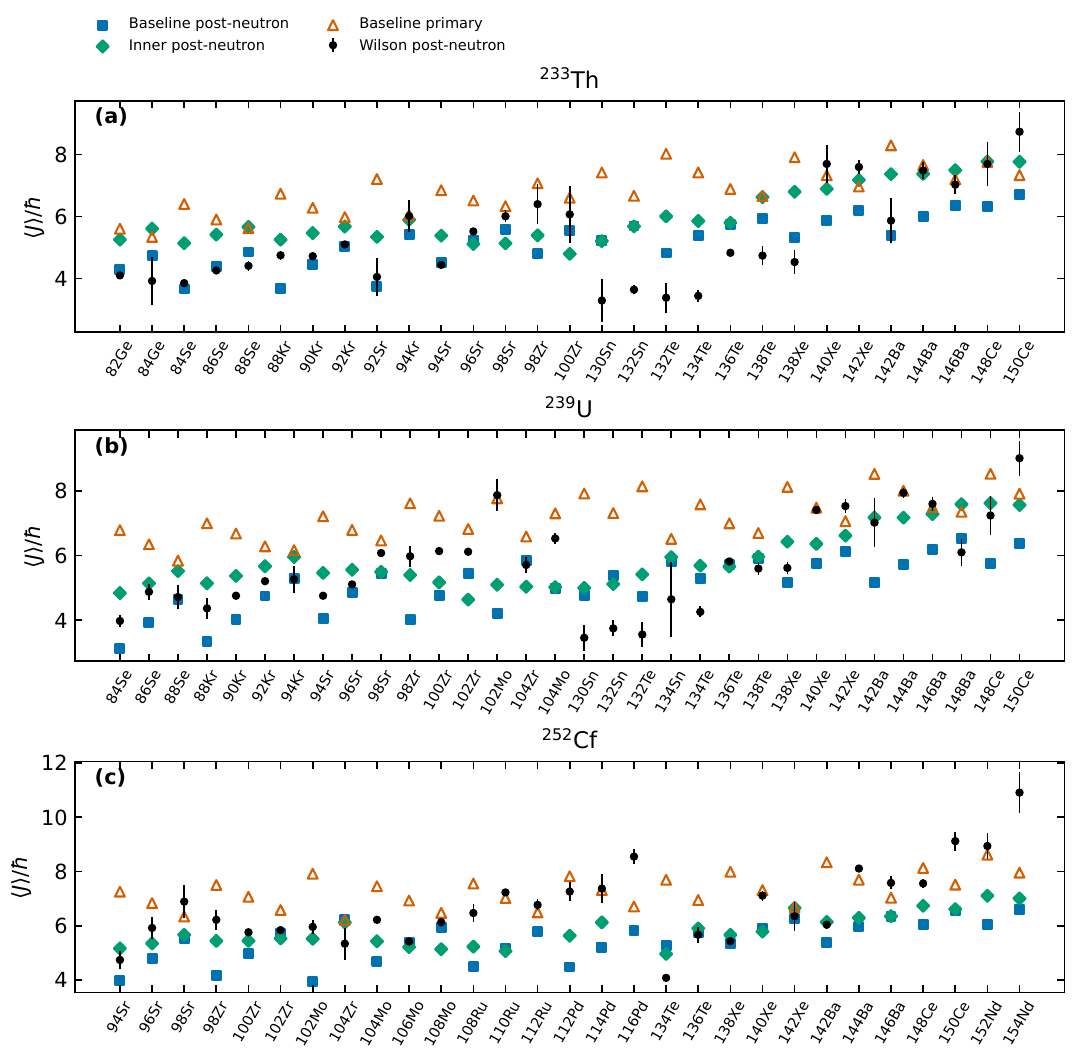}
\caption{Wilson residual-isotope means versus directly calculated primary spins followed by evaporation, illustrated with \(\alpha=1.23\) ℏ/neutron. Available refined residual estimates are shown separately. Primary means refer to the conditional ensemble feeding a residual isotope. Plotting code was prepared with assistance from ChatGPT (OpenAI); curves and markers are based on the stated numerical inputs.}
\end{figure}

No single decrement minimizes RMS across all geometries and systems. For ²⁵²Cf, the baseline is closer with 0.536, whereas 1.10 gives the smallest tested inner-branch RMS. The isotope counts differ, so these rows are not a fully matched comparison. For ²³⁹U, the inner geometry with 1.23 reduces RMS relative to baseline without establishing a universal decrement. The primary spin, excitation, and cascade must be treated jointly. No single directly measured table supplies both primary and residual spins: primary spins are independently calculated here, and the Wilson comparison concerns a reconstructed residual observable.

\section{Discussion}\label{discussion}

\subsection{Implications of weak experimental correlations}\label{implications-of-weak-experimental-correlations}

Figure 2(a) of Wilson et al.~\cite{ref2} shows the dependence of a selected fragment's mean residual spin on the partner's spectroscopic selection. The preceding numerical analysis extracted 25 vector markers and their errors for six isotopes in ²³⁸U(n,f). We verified the common-slope regression with an independent intercept per isotope, \(y_{ij}=a_i+bx_{ij}\). With diagonal weights \(1/\sigma_{ij}^2\), \(b=-0.01312\pm0.01287\). This conditional slope error does not include covariance between points. No significant linear dependence is indicated by this extraction.

\begin{figure}[htbp]
\centering
\includegraphics[width=0.98\linewidth,height=.65\textheight,keepaspectratio]{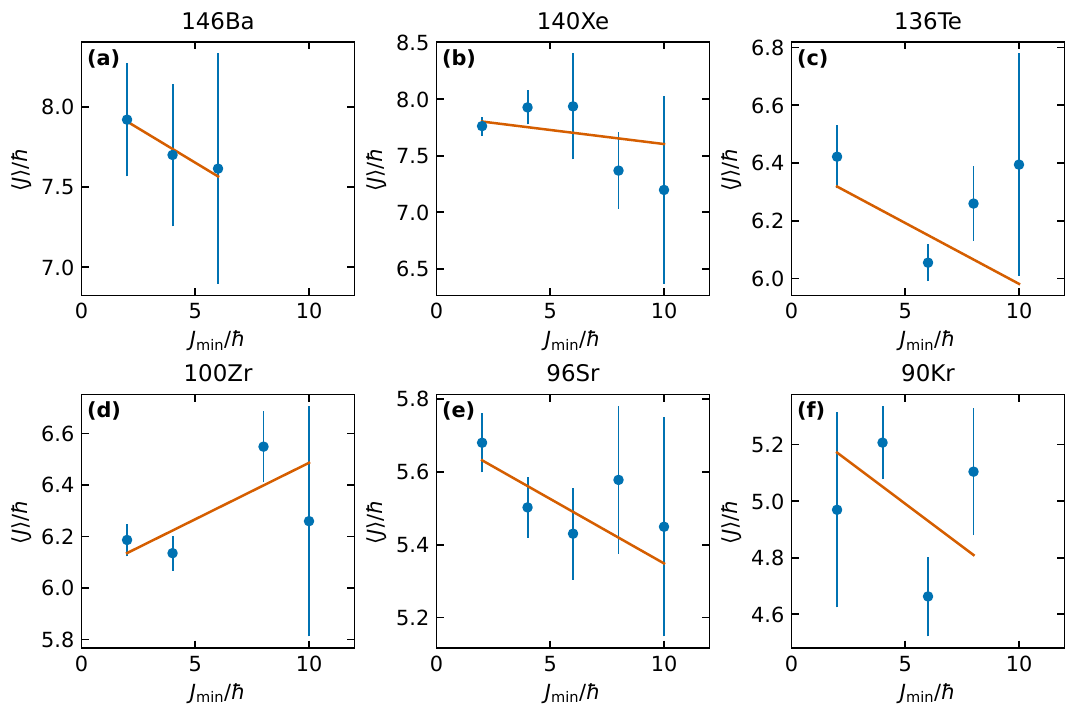}
\caption{Digitized points from Fig. 2(a) of Wilson et al.~\cite{ref2} and individual isotope regressions. Error bars reproduce the graphical intervals of the source. Plotting code was prepared with assistance from ChatGPT (OpenAI); curves and markers are based on the stated numerical inputs.}
\end{figure}

This slope is not a Pearson coefficient of primary spins: the abscissa is a spectroscopic gate, not the eventwise primary spin magnitude of the partner. Wilson's analysis uses information from γ transitions, whereas Eq. (16) describes the ground state before decay. Primary-spin coefficients cannot be compared directly without conditional detection probabilities, decay channels, full covariance, and event distributions. The small slope is compatible with the measured data, but does not establish small primary spatial correlations. Equations (20), (24), and (31) show how weak magnitude correlations can coexist with vector correlations.

\subsection{Stability of the orientation preference}\label{stability-of-the-orientation-preference}

For the 645 positively weighted baseline configurations, the preceding calculation carried out 1290 control integrations at separations shifted by ±0.2 fm, recomputing the Hessian on the control grid. The obtuse fraction changes by less than one percentage point even though absolute spins are sensitive to separation. To separate geometric effects from the change in the energy-selected channel population, the baseline and inner branches are also compared on exactly the same 452 channels.

\begin{table}[htbp]
\centering
\caption{Geometric comparison on matched support and with identical weights. The first three pairs of values give baseline and refined values; the obtuse fraction is in percent. Energies refer to the inner branch.}
\small
\begin{adjustbox}{max width=\linewidth}
\begin{tabular}{lccccc}
\toprule
System & \(\langle J_L\rangle\) & \(\langle J_H\rangle\) & \(P_{\mathrm{obt}}\) & \(\varepsilon_1\) (inner) & \(\varepsilon_2\) (inner) \\
\midrule
²³³Th & 6.232 → 7.066 & 7.437 → 8.471 & 43.00 → 43.03 & 1.480 & 2.731 \\
²³⁹U & 6.841 → 7.596 & 7.754 → 8.642 & 41.65 → 41.63 & 1.414 & 2.699 \\
²⁵²Cf & 7.265 → 7.732 & 8.020 → 8.565 & 41.00 → 40.98 & 1.254 & 2.401 \\
\bottomrule
\end{tabular}\end{adjustbox}
\end{table}

\begin{figure}[htbp]
\centering
\includegraphics[width=0.98\linewidth,height=.65\textheight,keepaspectratio]{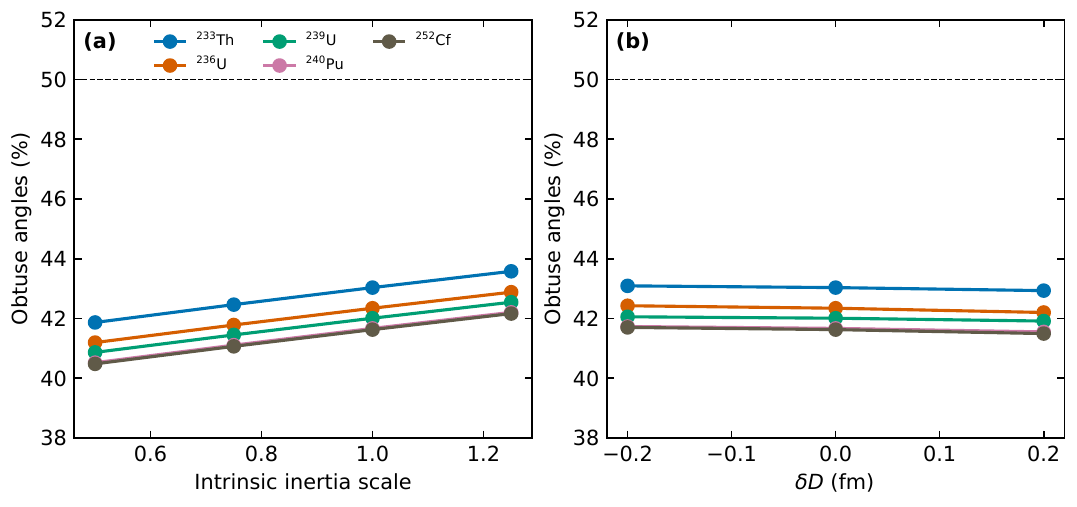}
\caption{Sensitivity of obtuse-angle fractions to center separation and common rescaling of intrinsic inertias at fixed potential. Geometry and inertia variations are evaluated separately. Plotting code was prepared with assistance from ChatGPT (OpenAI); curves and markers are based on the stated numerical inputs.}
\end{figure}

Table IX establishes that the almost unchanged asymmetry is not simply a consequence of channel reselection. On identical supports the geometric change noticeably affects the spins and energies but barely changes the mean difference of normalized stiffness and kinematic contributions. Within this range, the absolute spin scale and correlation direction have very different geometric sensitivities. This does not imply robustness to arbitrary changes of geometry or energy functional.

As a separate test, both intrinsic inertias were scaled by \(f=0.5,0.75,1,1.25\), holding \(I_0\) and \(K\) fixed. The baseline and inner geometries yield 5164 matrix solutions. In the baseline systems the obtuse fractions range from 40.48--41.87\% at \(f=0.5\) and 42.16--43.58\% at \(f=1.25\); the sign does not change. At \(f=0.5\), the mean heavy-fragment spin of ²⁴⁰Pu decreases to approximately 6.38 ℏ, still above the AMPLIFI value 4.48 ℏ. A common rescaling is not a Migdal or microscopic pairing calculation and does not probe separate variations of the two inertias. Changing the overall inertia scale alone does not resolve the discrepancies.

\subsection{The angular-momentum reservoir and competing couplings}\label{the-angular-momentum-reservoir-and-competing-couplings}

Randrup \cite{ref17} identified relative rotation as an angular-momentum reservoir in the statistical description because \(I_0\) is large compared with the intrinsic inertias. For vanishing total transverse angular momentum, the relative rotational energy is \((\mathbf J_1+\mathbf J_2)^2/(2I_0)\). A large \(I_0\) reduces the energy cost of compensation. This kinematic property also holds in the cold model, but does not imply independence of the quantum ground-state spins.

To quantify the comparison, define \(R_I=2I_0/(I_1+I_2)\) and retain the same matrices \(K,M\) and positive weights as in the baseline calculation. The scalar product \(\langle\mathbf J_1\cdot\mathbf J_2\rangle=2S_{12}\) is decomposed according to Eq. (15). Table X averages the individual channel contributions rather than multiplying averaged quantities. The width ratio is a ratio of means in the canonical convention of Eq. (10), not a ratio of mode energies or of contributions to individual spins.

\begin{table}[htbp]
\centering
\caption{Reservoir and covariance decomposition in the baseline mixtures. The contribution columns are weighted means in ℏ².}
\small
\begin{adjustbox}{max width=\linewidth}
\begin{tabular}{lcccccc}
\toprule
System & \(R_I\) & \(\bar C_w/\bar C_b\) & \(a_1a_2C_w\) & \(-C_b\) & \((a_2-a_1)C_{wb}\) & Sum \\
\midrule
²³³Th & 8.09 & 5.37 & 30.29 & -25.73 & 3.92 & 8.48 \\
²³⁶U & 8.22 & 5.44 & 30.34 & -25.44 & 4.08 & 8.98 \\
²³⁹U & 8.50 & 5.54 & 35.91 & -28.13 & 3.06 & 10.84 \\
²⁴⁰Pu & 8.61 & 5.55 & 37.23 & -28.65 & 2.69 & 11.28 \\
²⁵²Cf & 8.63 & 5.68 & 39.63 & -29.59 & 2.20 & 12.23 \\
\bottomrule
\end{tabular}\end{adjustbox}
\end{table}

The relative-motion inertia is 8.09--8.63 times the mean intrinsic inertia. The wriggling width exceeds the bending width by factors of 5.37--5.68, but after distributing the wriggling momentum between fragments its positive contribution is 30.29--39.63 ℏ², whereas the bending contribution is −25.44 to −29.59 ℏ². The positive mixed term of 2.20--4.08 ℏ² is not a contribution of a third independent mode. A much smaller positive sum remains. Weak magnitude correlations cannot be attributed solely to wriggling dominance: pure wriggling would align the two spin vectors. They reflect partial cancellation and the quadratic dependence in Eq. (20), with the mixture effect in Eq. (31) providing an additional contribution.

For a controlled comparison with the statistical equilibrium model \cite{ref17}, take the classical canonical density for the transverse Hamiltonian (12) at a common temperature \(T_{\mathrm{cl}}\) in energy units. The potential factor separates upon integration over the angles, leaving a momentum covariance independent of \(K\). Inverting Eq. (9) gives

\begin{equation}
M^{-1}=\operatorname{diag}(I_1^{-1},I_2^{-1})+\frac{\mathbf 1\mathbf 1^{\mathsf T}}{I_0},\qquad S_{\mathrm{cl}}=T_{\mathrm{cl}}M.\tag{34}
\end{equation}

Thus the correlation of like transverse components is

\begin{equation}
\rho_{\mathrm{cl}}=-\sqrt{\frac{I_1I_2}{(I_0+I_1)(I_0+I_2)}}=-m.\tag{35}
\end{equation}

Equation (35) describes planar canonical equilibrium with zero total angular momentum and equal mode temperatures; it is not a new FREYA simulation. The common temperature cancels from the correlation and angle density, so its absolute value is not specified. Only the kinematic anticorrelation remains in this limit, and it vanishes as \(I_0\to\infty\). The cold ground-state limiting covariance is different:

\begin{equation}
S_{\infty}=\frac{K+d_{\infty}\operatorname{diag}(I_1,I_2)}{2\sqrt{\operatorname{tr}[\operatorname{diag}(I_1,I_2)^{-1}K]+2d_{\infty}}},\quad d_{\infty}=\sqrt{\frac{\det K}{I_1 I_2}}.\tag{36}
\end{equation}

Here \(K,I_1,I_2\) are held fixed. A finite positive \(K_{12}\) leaves a positive correlation. This is a formal inertia substitution, not an increase of the physical separation \(D\), which would also change the potential. The two parameters need not be independent during actual separation.

\begin{table}[htbp]
\centering
\caption{Cold, classical, and limiting cold calculations on identical baseline channels and weights. Subscripts 0, cl, and ∞ refer to the three respective calculations.}
\small
\begin{adjustbox}{max width=\linewidth}
\begin{tabular}{lcccccc}
\toprule
System & \(\bar\rho_0\) & \(\bar\rho_{\mathrm{cl}}\) & \(\bar\rho_{\infty}\) & \(\bar P_{>,0}\), \% & \(\bar P_{>,\mathrm{cl}}\), \% & \(\bar P_{>,\infty}\), \% \\
\midrule
²³³Th & +0.1393 & -0.1032 & +0.1885 & 43.03 & 55.16 & 40.58 \\
²³⁶U & +0.1531 & -0.1015 & +0.2015 & 42.34 & 55.07 & 39.93 \\
²³⁹U & +0.1598 & -0.1013 & +0.2079 & 42.01 & 55.06 & 39.61 \\
²⁴⁰Pu & +0.1667 & -0.1011 & +0.2148 & 41.67 & 55.06 & 39.26 \\
²⁵²Cf & +0.1675 & -0.1011 & +0.2156 & 41.63 & 55.06 & 39.22 \\
\bottomrule
\end{tabular}\end{adjustbox}
\end{table}

\begin{figure}[htbp]
\centering
\includegraphics[width=0.98\linewidth,height=.65\textheight,keepaspectratio]{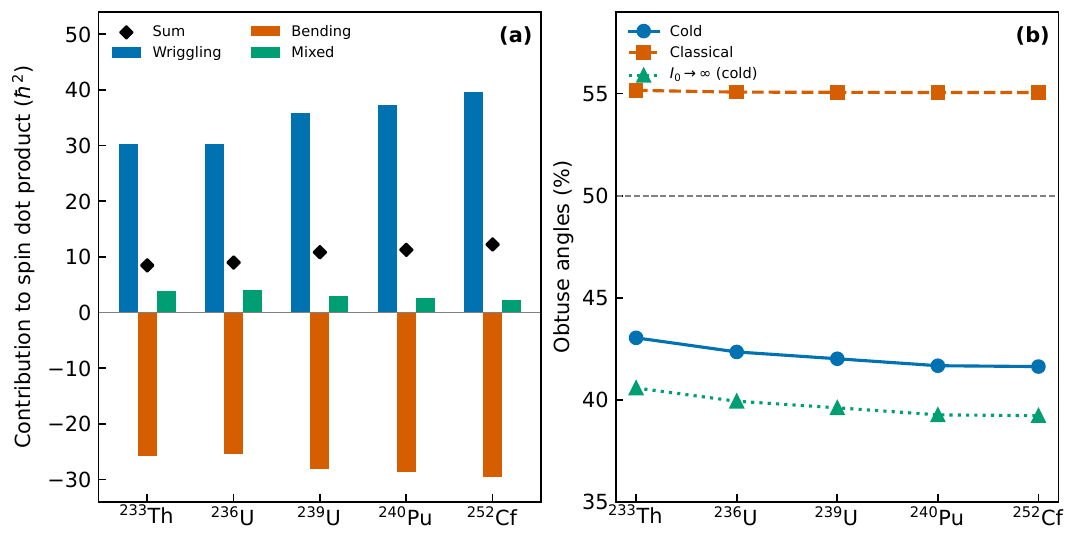}
\caption{(a) Mean wriggling, bending, and mixed contributions to the spin scalar product and their sum. (b) Obtuse-angle fractions for the cold ground state, classical canonical case, and the formal infinite-relative-inertia limit at fixed potential. Plotting code was prepared with assistance from ChatGPT (OpenAI); curves and markers are based on the stated numerical inputs.}
\end{figure}

At identical inertias, the signs are opposite: \(\bar\rho_0=0.1393\)--0.1675 and \(\bar\rho_{\mathrm{cl}}=-0.1032\) to −0.1011. The classical obtuse fractions are 55.05--55.16\%, while the cold fractions are 41.63--43.03\%. Mean within-channel magnitude correlations remain small in both cases (0.0191--0.0273 and 0.0094--0.0098, respectively). In contrast to classical equilibrium, the formal infinite-inertia cold correlations increase to 0.1885--0.2156 and obtuse fractions decrease to 39.22--40.58\%. All 839 baseline and 452 alternative solutions were evaluated. The maximum independent matrix discrepancy is \(8.53\times10^{-14}\) ℏ² and the decomposition error in Eq. (15) is \(2.49\times10^{-14}\) ℏ². Numerical densities are normalized within \(10^{-10}\) and integrated obtuse fractions agree with Eq. (24) within \(10^{-7}\).

The reservoir interpretation is thereby refined: a large \(I_0\) weakens kinematic coupling, but does not eliminate potential coupling in the quantum ground state. Spin orientations distinguish the two limits much more directly than the small magnitude correlations. This is neither a refutation of Randrup's statistical model \cite{ref17} nor a determination of the temperature of a real fissioning nucleus; it compares two specified reductions at identical inertias. Independently constrained stiffnesses, matched quantum projections, and measurements of spatial correlations are required.

\subsection{Physical limits of the cold model and further tests}\label{physical-limits-of-the-cold-model-and-further-tests}

The correlation sign in the cold model reflects competing potential and kinematic terms in Eq. (28). Positive correlations are not an inevitable consequence of absent twisting excitation: \(k<m\) would favor obtuse angles in the same planar cold model. The orientation thus tests a specified angular Hamiltonian rather than directly measuring the temperature of the fissioning nucleus.

Possible reasons for the discrepancies include the local interaction approximation, rigid inertias, omitted nonaxial shapes and intrinsic longitudinal projections, initial compound spin, and subsequent separation dynamics. No single explanation is established here. The longitudinal projections in Ref. \cite{ref6} constrain a strictly transverse model but do not themselves establish thermal twisting. Recent calculations \cite{ref18,ref19} address other mechanisms of spin generation and correlation; the excitation and the fissioning systems must be matched before numerical comparison, in particular for hot systems in Ref. \cite{ref18}. Quantum angular motion and energy selection are also explored in Refs. \cite{ref14,ref15,ref21}. These studies preclude a claim of priority for using an angular potential but motivate evaluation of its complete Hessian.

Further tests require angular derivatives of a more complete energy functional and structure-dependent inertias, followed by joint angular-momentum projection with allowed compound-spin projections. Deformation, intrinsic, radial kinetic, and zero-point energies should be included separately; actual coincidence yields and neutron allocations should then be generated. An eventwise cascade would permit comparison with the same spectroscopic observables as Wilson. The testable result here is the relationship between normalized matrix elements and the orientation asymmetry, not a universal 42\% for cold configurations. Persistent discrepancies with matched quantum observables and independently determined \(K,M\) would require revising the prescission state or including postscission dynamics.

\section{Conclusions}\label{conclusions}

We have calculated primary spins and their spatial correlations for five fissioning systems within the cold axial model. Unlike Ref. \cite{ref1}, bending and wriggling energies are obtained from angular derivatives of the effective potential and inertias of reconstructed shapes, rather than prescribed as constants. The baseline eigenenergies are 1.10--1.17 and 2.01--2.14 MeV; the lower mode is predominantly bending. Two fixed uncoupled energies with the new inertias do not reproduce the coupled calculation.

The sign criterion in terms of normalized mixed stiffness and kinematic coupling is confirmed for all 1291 stable baseline and alternative solutions. Their baseline mixtures favor acute angles, with 41.6--43.0\% obtuse angles. FREYA and the more complete projection of Scamps et al.~slightly favor obtuse angles. Agreement in weak magnitude correlations therefore does not imply agreement in spin orientation; the distinction persists under the considered geometry and inertia variations.

At identical inertias, statistical equilibrium and the cold ground state produce opposite correlation signs. Even in the formal limit \(I_0\to\infty\) at fixed \(K\), the potential coupling survives in the cold state. Partial cancellation of wriggling and bending contributions helps explain why both descriptions can yield small magnitude correlations despite different spin orientations. This extends the reservoir interpretation of Randrup \cite{ref17} to the specified cold model without claiming priority for the reservoir concept itself.

The magnitude coefficient is quadratic in weak component correlations, whereas the orientation asymmetry is linear. Within-channel covariance can also cancel between-channel covariance, yielding a nearly zero mixture Pearson coefficient for ²³⁶U despite nonuniform spin orientations. Weak magnitude correlations alone cannot establish independent spin generation or a postscission origin.

AMPLIFI comparisons reproduce average light spins more closely than individual means, particularly for heavy ²⁴⁰Pu fragments. Comparisons with Wilson residual spins depend on geometry and evaporation. The small extracted spectroscopic slope does not determine primary Pearson correlations. The proposed method specifies a physical condition for spin orientation and enables a joint test of the cold model through several observables; it does not remove the need for refined interactions, inertias, and decay cascades.

This study isolates the cold ground-state correlation-sign criterion and the fixed-inertia reservoir comparison. Two companion manuscripts using the same archived matrices develop different tests: the finite-occupation crossover and the orbital-conditioned spin diagnostic. The common inputs and baseline results are not independent replications.

\textbf{Data availability.} The accompanying computational material contains numerical inputs, channel and mixture distributions, tests of Eqs. (25)--(28), sensitivity calculations, and figures. The Python scripts reproduce the calculations and reservoir comparison. A newly prepared Wolfram Language implementation is supplied for independent verification; it has not been executed in a licensed Mathematica kernel. Primary Wilson events and their full covariance matrices are unavailable.

\begin{acknowledgments}
This work was supported by the Russian Science Foundation under Grant No. \href{https://www.rscf.ru/upload/iblock/baf/99vu3fqix0zcid15a6ducl73rrh3gyp3.pdf}{25-22-00697}.
\end{acknowledgments}
\section*{AI Use Disclosure}
\href{https://chatgpt.com/}{ChatGPT} (OpenAI; GPT-6.2 was used for final preparation) was used to assist with mathematical derivations and checks, numerical analysis, Python and Wolfram Language code development, preparation of data-derived plots, literature organization, English translation, and language editing. The authors have reviewed the manuscript and take full responsibility for its content, including the scientific claims and conclusions.

\end{document}